\documentclass[runningheads]{llncs}
\usepackage{graphicx}
\usepackage{diagbox}
\begin{document}
\title{Diagnosis of Helicobacter pylori using AutoEncoders for the Detection of Anomalous Staining Patterns in Immunohistochemistry Images}
\titlerunning{Diagnosis of Helicobacter pylori using AutoEncoders}
%
\author{Pau Cano\inst{1,2} \and
Álvaro Caravaca \and
Debora Gil\inst{1,2}\and
Eva Musulen\inst{3,4}}

\authorrunning{Cano, P et al.}
%
%
\institute{Comp. Sci. Dep., Universitat Autònoma de Barcelona \and
Computer Vision Center, campus UAB, Barcelona, Spain
\email{\{debora,pcano\}@cvc.uab.cat} \and
Pathology Dep.,Hospital Universitari General de Catalunya \and Josep Carreras Institute, Barcelona, Spain
\email{eva.musulen@quironsalud.es,emusulen@carrerasresearch.org}}
\maketitle              
\begin{abstract}

This work addresses the detection of Helicobacter pylori a bacterium classified since 1994 as class 1 carcinogen to humans. By its highest specificity and sensitivity, the preferred diagnosis technique is the analysis of histological images with immunohistochemical staining, a process in which certain stained antibodies bind to antigens of the biological element of interest. This analysis is a time demanding task, which is currently done by an expert pathologist that visually inspects the digitized samples.

We propose to use autoencoders to learn latent patterns of healthy tissue and detect \textit{H. pylori} as an anomaly in image staining. Unlike existing classification approaches, an autoencoder is able to learn patterns in an unsupervised manner (without the need of image annotations) with high performance. In particular, our model has an overall 91\% of accuracy with 86\% sensitivity, 96\% specificity and 0.97 AUC in the detection of \textit{H. pylori}. 

\end{abstract}
\section{Introduction}

The bacterium \textit{Helicobacter pylori} (H. pylori) is the main cause of gastritis, an inflammation of the gastric mucosa that can lead to other serious diseases, such as gastric ulcer and even cancer. Early detection of this bacterium is essential for the effective diagnosis and treatment of these pathologies. In addition, studies show that more than 50\% of the world's population has been infected by the bacterium, with a prevalence that exceeds 80\% in adults over fifty \cite{yang}. 

The diagnosis of \textit{H. pylori} is usually made by conventional histology on gastric biopsies using different techniques for staining tissue samples. Usual stainings include the generic hematoxylin and eosin (H\&E) and more specific stains such as Giemsa, Warthin-Starry silver (W-S), Genta or immunohistochemical 
staining. Among them, the more specific one is immunohistochemical 
staining \cite{Batts}. This technique allows the visualization of the bacterium through the staining of specific proteins present in its membrane. In this manner, \textit{H. pylori} stains with a color different from the one of other tissue, which avoids false detection of \textit{H. pylori} due to other gram-negative bacteria present in the sample. Immunohistochemical staining gives the specific protein of \textit{H. pylori} a reddish hue, while other tissue remains in a blue hue. Although this facilitates the visual identification of \textit{H. pylori}, a pathologist must carefully inspect the whole immunohistochemistry images in order to identify areas with \textit{H. pylori}. Since the bacteria are only located at the borders of the tissue samples, the pathologists must carefully inspect a zoom-up area for all points belonging to the border. Given the huge size of images (120000x16000 pixels) and the fact that several tissue samples can be in the same image, this manual inspection is a highly time consuming task that becomes harder the lower the concentration
of \textit{H. pylori} is.  

\begin{figure}[!h]
\centering
	\includegraphics[width=0.9\textwidth]{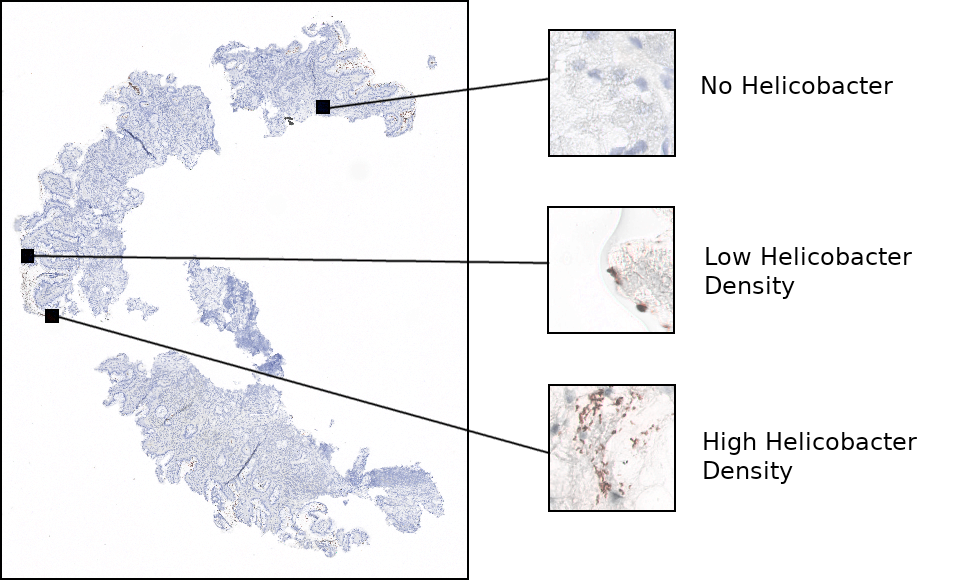}
	\caption{Left: Histology sample with immunohistochemical staining. Right: 3 windows of the same histological sample showing different levels of \textit{Helicobacter pylori} density}
	\label{fig-histology-inmuno}
\end{figure}

Figure \ref{fig-histology-inmuno} shows an immunohistochemical image with presence of \textit{H. pylori} in a sample and three close-up of tissue border regions with different density (negative, low and high) of the bacteria on the window images shown on the the right side of the figure. While the window with high presence of H. pylori is easily identified, the window with low density needs a more careful inspection in order to detect the reddish spots of H. pylori and avoid confusion with other artifacts that can be in the sample.


Due to the recent digitalization of
histopathological images, there is a lack of artificial intelligence
methods for their analysis. In this work, we propose a method to automatically analyze an image of a histological sample of gastric tissue that
has been immunohistochemically stained for the detection of
\textit{H. pylori}.

\subsection{State-of-the-Art}
\label{SoA}

Although Deep Learning, DL (and other Artificial Intelligence) models have demonstrated good performance on several histopathologic tasks \cite{Salto}, there are not many works addressing the detection of \textit{H. pylori}. Existing works \cite{Liscia,Klein,Zhou} are DL methods based on convolutional neural networks for the classification of cropped images extracted from tissue samples into \textit{H. pylori} positive and negative samples. 

In \cite{Klein} the authors trained a compact VGG-style architecture on, both, Giemsa and H\&E slides. The trained network was used to highlight regions of \textit{H. pylori} presence and tested as decision support system to pathologists. The network was able to classify Giemsa stained samples with a sensitivity of 1 with a low specificity of 0.66. In \cite{Liscia} the authors also use a model similar to \cite{Klein} but trained on silver staining samples. The performance was also tested in cropped patches achieving a sensitivity and specificity of, respectively, $0.89$ and $0.87$ at the cost of a significant amount of false positives having only $77\%$ of precision in the detection of patches with \textit{H. pylori}. 

In \cite{Zhou}, the authors proposed an ensemble model of the output probabilities of 3 ResNet-18 and 3 DenseNet-21 models trained on patches cropped from H\&E-stained Whole-Slide Images (WSI). Patch-level probabilities were aggregated into WSI-level probabilities by averaging the top 10 patch-level probabilities from a each section. The ensemble achieved a sensitivity of $0.87$, a specificity of $0.92$ and F1-score of 0.89 for the diagnosis of WSI. The model was also tested as DL support system to improve the performance of a pathologist, who improved the accuracy and performance when diagnosing \textit{H. pylori} positive samples, but resulted in higher uncertainty when diagnosing \textit{H. pylori} negative samples. 

As far as we know there are no works addressing the diagnosis of immunohistochemically stained WSI. One of the main challenges for the use of classification approaches for the identification of \textit{H. pylori} in histological images is the collection of enough annotated data, since this implies a time consuming visual inspection and identification of patches containing the bacteria. 

In this work we pose the detection of \textit{H. pylori} as a detection of anomalies in the staining of tissue by means of an autoencoder able to learn patterns of non-infected tissue without the need of annotated data. An autoencoder is a type of neural network with an encoder-decoder architecture that learns a latent representation space of the input data. The encoder transforms the input data into a lower-dimensional representation called latent code, while the decoder   reconstructs the original data from this latent space. The latent space is learned to minimize the reconstruction mean square error between the original input image and the one reconstructed by the decoder and, thus, it can be trained in an unsupervised fashion. 

By training the autoencoer with patches (windows) extracted from patients without \textit{H. pylori}, the latent space is a representation of non-infected tissue and, thus, windows with the presence of \textit{H. pylori} are poorly reconstructed. A function of this reconstruction error in HSV color space allows the detection of windows with \textit{H. pylori} and a final diagnosis of the WSI by the aggregation of the diagnosis of windows extracted from the tissue borders. 

\section{Detection of \textit{H. pylori} using Autoencoders}

\begin{figure}[!h]
\centering
	\includegraphics[width=0.9\textwidth]{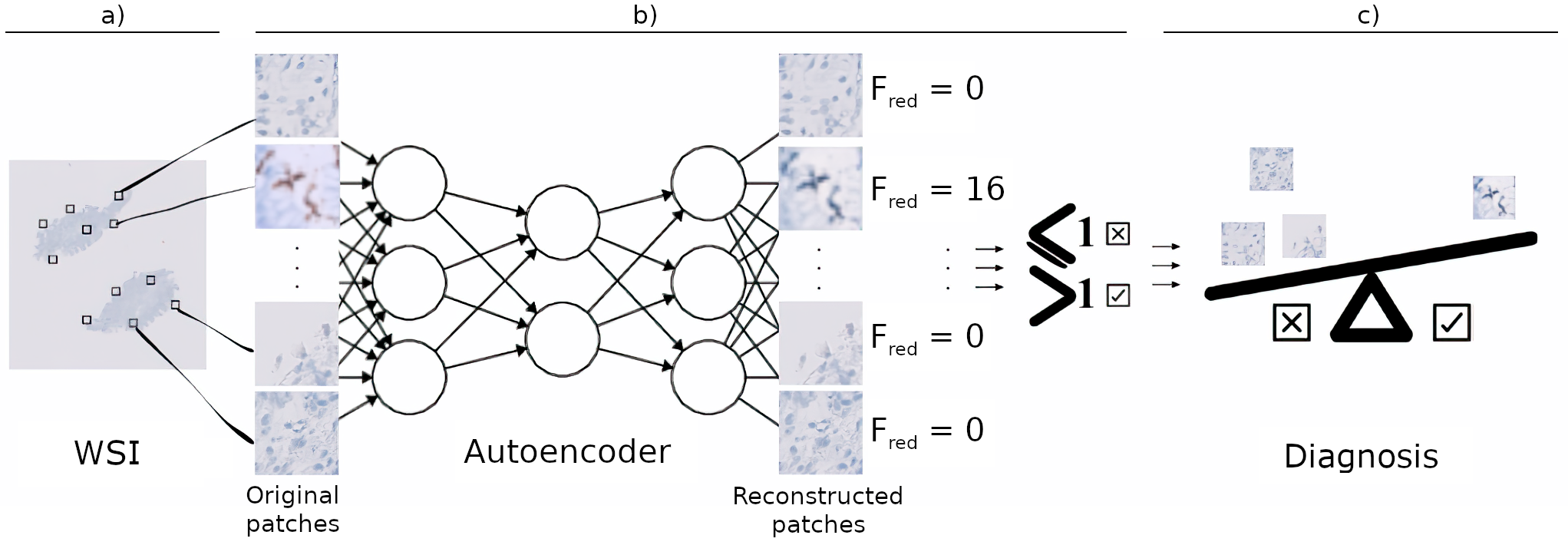}
	\caption{Schema of the main steps in the detection of H. pylori}
	\label{fig-autoencoder-architecture}
\end{figure}

Our method has the following steps (sketched in  figure \ref{fig-autoencoder-architecture}): detection of areas of interest in the image, detection of anomalous stained elements in each region of interest, and aggregation of each region of interest in the image for the diagnosis of the sample. Since \textit{H. pylori} is located along the
border, first a series of contour detections around an automatically detected mask are used to detect the borders
of the tissue sample. Patches are defined by sliding windows of size 256x256 pixels cropped along pixels belonging to such borders. This set of windows are the input to the autoencoder for their classification into positive (there is \textit{H. pylori} presence in the window) or negative (there is not \textit{H. pylori} in the window) cases, using a metric based on the loss of red-like pixels in reconstructions. Finally, the percentage of positive windows defines a probability for the final classification of the sample.

\begin{figure}[!h]
\centering
\begin{tabular}{c|c}
  \includegraphics[width=0.4\textwidth]{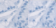}   & \includegraphics[width=0.4\textwidth]{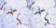} \\
     (a) & (b)
\end{tabular}
	\caption{Reconstructions of a healthy, (a), and infected, (b), windows. For each sub-figure, left images are the original inputs and rigth images, the reconstructions.}
	\label{fig-reconstruction-healthy}
\end{figure}

The autoencoder is trained with windows extracted from patients without \textit{H. pylori} for learning a representation space of normality (non-infected tissue). The proposed autoencoder has 3 convolutional blocks with one convolutional layer, batch normalization and leakyrelu activation each. The size of the convolutional kernel is 3 and the number of neurons and stride of each layer are, respectively, [32,64,64] and [1,2,2]. Figure \ref{fig-reconstruction-healthy} shows the difference in the reconstructions of a non-infected window, fig.\ref{fig-reconstruction-healthy}(a), and a window with \textit{H. pylori}, fig.\ref{fig-reconstruction-healthy}(b). The reconstruction of the healthy window looks like the original input image,  while the autoencoder has modified the coloration of the tissue in the reconstruction of the window with \textit{H. pylori}.  In particular, the reconstruction has a color conversion to the blue hue and has lost the red-like areas associated with the presence of \textit{H. pylori}. We use this difference in reconstructions to detect the presence of \textit{H. pylori} as follows. 

The presence of \textit{H. pylori} in a window is computed using the fraction of red-like pixels, labelled $F_{red}$, lost between original and reconstructed images. If $F_{red}>1$, it indicates a loss of red-like pixels, and the window is labelled as having \textit{H. pylori}. The red-like pixels are computed applying a filter in HSV color space. In this color space, pixels with presence of \textit{H. pylori} have a hue in the range $[-20,20]$ and, thus, the area of red-like pixels is given by the number of pixels with hue in $[-20,20]$. 

The percentage of patches in an histological image with $F_{red}>1$ defines the probability of presence of \textit{H. pylori} in the sample. The optimal threshold of this probability is obtained from the ROC curve as the probability of the closest point to $(0,1)$. Samples with a percentage of positive patches above this threshold are diagnosed as \textit{H. pylori} positive.

\section{Experiments}

Our method was tested on our own database from the Department of Pathology of the Hospital Universitari General de Catalunya-Grupo Quironsalud. 
The database consisted of 245 gastric biopsies scored by a pathologist according to \textit{H. pylori} density as NEGATIVE (Healthy), LOW DENSITY and HIGH DENSITY. Of the 245 patients included in the study, 117 (47.8\% of the total) are classified as NEGATIVE, while 128 are classified as POSITIVE (LOW and HIGH DENSITY) with presence of \textit{H. pylori}. 


Biopsies from the Department of Pathology of the Hospital Universitari General de Catalunya-Grupo Quironsalud of antral or body gastric mucosa were used. Formalin-fixed, paraffin-embedded tissue sections were analyzed using standard IHC techniques: immunostaining was performed automatically using a Ventana BenchMark ULTRA machine (\textit{Roche, Basel, Switzerland}) using the monoclonal primary antibody anti-Hp (\textit{clone SP48, Ventana Medical Systems, Inc., 1910 E. Innovation Park Drive, Tucson, Arizona 85755 USA}). An external positive control was included on each slide. All stained slides were scanned with an Ultra-Fast 180 slide scanner provided by Philips (\textit{Philips IntelliSite Pathology Solution}) to obtain WSI.

Each image has 3 WSI containing several tissue samples each, of which two are used for the pathological diagnosis, and the third one is a quality control slide. We have used the first diagnostic slide of the healthy cases to train the autoencoder and the second one of all patients to test the performance of the system in the diagnosis of \textit{H. pylori}. For each healthy patient, 50 windows where randomly cropped from tissue borders of the first sample slide, which gives a total number of 5850 windows for training models. For the sake of a higher computational speed, windows were resize from $224\times 224$ pixels to  $28\times 28$. 

The performance metrics we have considered are the precision, recall and F-1 score for each diagnostic class (positive \textit{H. pylori} or negative \textit{H. pylori}). In order to allow for statistical assessment of the performance, the test set was split in 10 folds stratified by patient. For each fold, the optimal cutting point of the ROC curve was calculated from the training fold and tested in the independent set of patients.

Table \ref{tab:resultats-autoencoder} reports statistical summaries (average $\pm$ standard deviation) for the quality metrics. The proposed system has a optimal average specificity of 0.96 with a good average sensitivity of 0.86, which yields a F1-score of 0.91 and accuracy of 0.91 for the detection of \textit{H. pylori}. Comparing to existing methods using other staining, we achieve a higher specificity with similar sensitivity. 
Table \ref{tab:confusion-matrix-autoencoder} shows the confusion matrix of the samples' diagnosis. Of 245 patients, only 23 have been incorrectly classified. 

\begin{table}[h]
\caption{Statistical Summary of the 10-fold Validation}
\label{tab:resultats-autoencoder}
\begin{center}
\begin{tabular}{|c|c|c|c|}
  \hline
  & negative \textit{H. pylori} & positive \textit{H. pylori} & Average \\
  \hline
  Precision & 0.86 $\pm$ 0.1 & 0.96 $\pm$ 0.07 & 0.91\\
  \hline
  Recall & 0.96 $\pm$ 0.09 & 0.86 $\pm$ 0.13 & 0.91 \\
  \hline
  f-1 score & 0.91 $\pm$ 0.06& 0.90 $\pm$ 0.07 & 0.91 \\
  \hline
\end{tabular}%
\end{center}
\end{table}

\begin{table}[h]

\caption{Confusion matrix of the samples' diagnosis and the diagnosis predicted by the autoencoder}
\label{tab:confusion-matrix-autoencoder}
\begin{center}

\begin{tabular}{|c|c|c|}
  \hline
  \diagbox{Ground Truth}{Predicted} & \textit{H. pylori} & No \textit{H. pylori} \\
  \hline
  \textit{H. pylori} & 110 (TP) & 18 (FP) \\
  \hline
  No \textit{H. pylori} & 5 (FN) & 112 (TN) \\
  \hline
\end{tabular}%
\end{center}
\end{table}

\begin{figure}[!h]
\centering
	\includegraphics[width=0.85\textwidth]{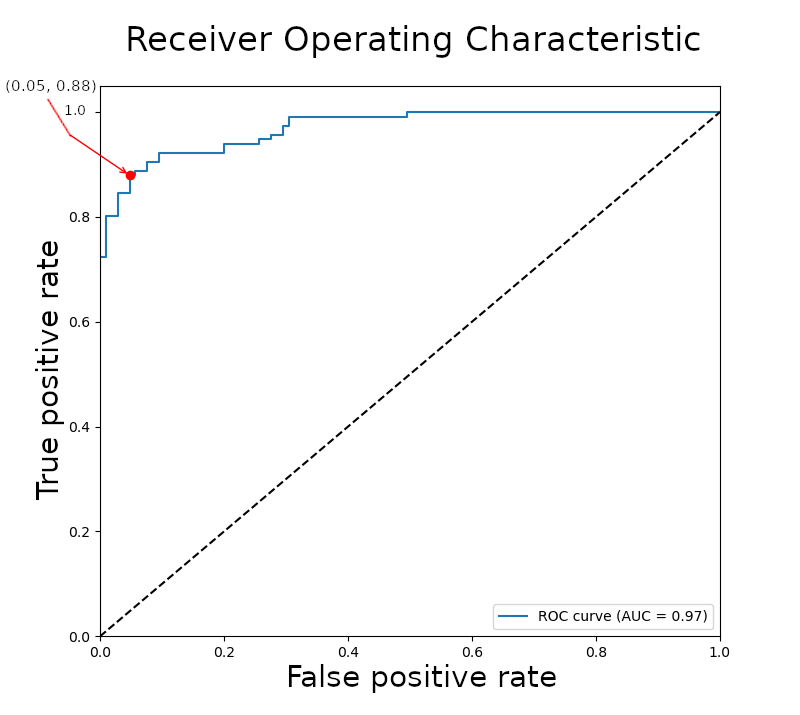}
	\caption{ROC curve averaged for the 10 folds}
	\label{fig-roc-autoencoder}
 
\end{figure}

Figure \ref{fig-roc-autoencoder} shows the ROC curve averaged for the 10 folds with the point defining the optimal threshold highlighted in red. It is noticeable the stability of this cutting point across folds, with a variability of 0.1 in the ranges (6.18\% $\pm$ 0.10\%) of the thresholding value of the probability of H. pylori. Additionally, the ROC curves have an average AUC of 0.961, which is superior to the ones achieved by other systems mentioned in section \ref{SoA}.

\begin{figure}[!h]
\centering
	\includegraphics[width=0.85\textwidth]{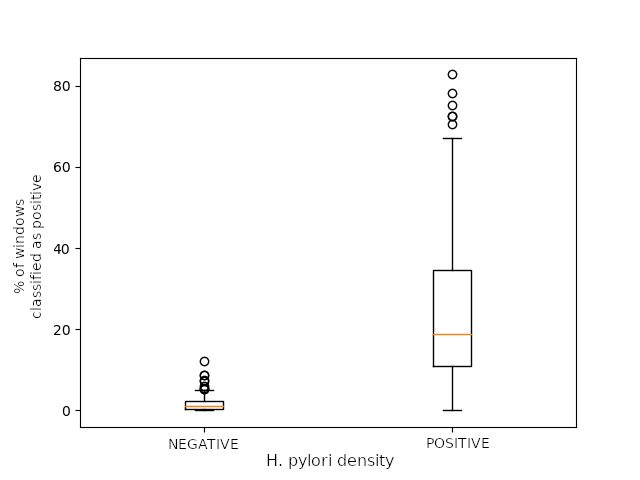}
	\caption{Boxplot of the percentage of windows detected as positive by the autoencoder for the two diagnosis}
	\label{fig-boxplot-autoencoder}
\end{figure}

Finally, figure \ref{fig-boxplot-autoencoder} shows boxplots for the percentage of positive windows detected by the autoencoder for POSTIVE and NEGATIVE diagnosis. There is a substantial difference between the two distributions. In particular, for NEGATIVE cases, the percentage of windows detected as positive is in most cases under 5\% with only some outliers, which explains the high specificity of out approach. 


\section{Conclusions}
We have presented a first DL system for the diagnosis of \textit{H. pylori} on immunohistochemically stained samples based on autoencoders trained to obtain a normality pattern from non-infected samples. Autoencoders are able to detect \textit{H. pylori} as an anomaly in staining in a self-learning approach that does not require annotation of image patches. This is a main advantage over existing classification approaches working with other kinds of staining and yields higher specificity (0.96 vs 0.92) with similar sensitivity, which is a clinical requirement to avoid unnecessary treatments.  


Additionally, modifying by a small value the threshold in the system that separates between \textit{H. pylori} positive and \textit{H. pylori} negative cases based on the percentage of windows detected with the bacterium would allow for increased precision without affecting much the recall of the system, or viceversa. 

%
%

%
%
%
\bibliographystyle{splncs04}
\bibliography{references}

\begin{thebibliography}{1}
\providecommand{\url}[1]{\texttt{#1}}
\providecommand{\urlprefix}{URL }
\providecommand{\doi}[1]{https://doi.org/#1}

\bibitem{Batts}
Batts, K., Ketover, S., Kakar, S., et~al: Gastrointestinal pathology society.
  appropriate use of special stains for identifying helicobacter pylori:
  Recommendations from the rodger c. haggitt gastrointestinal pathology
  society. Am J Surg Pathol  \textbf{37}(11),  e12--22 (2013)

\bibitem{Klein}
Klein, S., Gildenblat, J., Ihle, M.A., at~al: Deep learning for sensitive
  detection of helicobacter pylori in gastric biopsies. BMC gastroenterology
  \textbf{20}(1),  1--11 (2020)

\bibitem{Liscia}
Liscia, D.S., D’Andrea, M., Biletta, E., et~al: Use of digital pathology and
  artificial intelligence for the diagnosis of helicobacter pylori in gastric
  biopsies. Pathologica  \textbf{114}(4), ~295 (2022)

\bibitem{Salto}
Salto-Tellez, M., Maxwell, P., Hamilton, P.: Artificial intelligence-the third
  revolution in pathology. Histopathology  \textbf{74}(3),  372--376 (2019)

\bibitem{yang}
Yang, J.: Treatment of helicobacter pylori infection: Current status and future
  concepts. World J Gastroenterol  \textbf{20(18)},  5283--93 (2014)

\bibitem{Zhou}
Zhou, S., Marklund, H., Blaha, O., et~al: Deep learning assistance for the
  histopathologic diagnosis of helicobacter pylori. Intelligence-Based Medicine
   \textbf{1},  100004 (2020)

\end{thebibliography}
%







\end{document}